\documentclass[12pt,prd]{article}
\pdfoutput=1
\usepackage{jheppub}
\usepackage{rotating}
\usepackage{array}
\usepackage{amsmath}
\usepackage[normalem]{ulem}
\usepackage{slashed}
\usepackage{booktabs}
\usepackage[pdftex,table]{xcolor}
\usepackage{units}
\usepackage{xfrac}
\usepackage{mathtools}
\usepackage{empheq}
\usepackage[]{units}
\usepackage{multirow}
\usepackage{amssymb}
\usepackage{url}
\usepackage{comment}
\usepackage{paralist}

\usepackage{tikz}

\def\beq{\begin{equation}}
\def\eeq{\end{equation}}
\newcommand{\bea}{\begin{eqnarray}\begin{aligned}}
\newcommand{\eea}{\end{aligned}\end{eqnarray}}
\def\bitem{\begin{itemize}}
\def\eitem{\end{itemize}}

\definecolor{darkpurple}{rgb}{0.5, 0.2, 0.8}
\definecolor{darkblue}{rgb}{0.0, 0.0, 0.8}
\definecolor{darkgreen}{rgb}{0.0, 0.4, 0.0}
\definecolor{darkred}{rgb}{0.5, 0.0, 0.0}
\usepackage{hyperref}
\hypersetup{
    linktocpage,
     colorlinks,
     citecolor=darkgreen,
     linkcolor= darkgreen,
     urlcolor=darkgreen
}

\abstract{
A growing number of weak- and unsupervised machine learning approaches to anomaly detection are being proposed to significantly extend the search program at the Large Hadron Collider and elsewhere.   One of the prototypical examples for these methods is the search for resonant new physics, where a bump hunt can be performed in an invariant mass spectrum.   A significant challenge to methods that rely entirely on data is that they are susceptible to sculpting artificial bumps from the dependence of the machine learning classifier on the invariant mass.  We explore two solutions to this challenge by minimally incorporating simulation into the learning.  In particular, we study the robustness of Simulation Assisted Likelihood-free Anomaly Detection (SALAD) to correlations between the classifier and the invariant mass.  Next, we propose a new approach that only uses the simulation for decorrelation but the Classification without Labels (CWoLa) approach for achieving signal sensitivity.  Both methods are compared using a full background fit analysis on simulated data from the LHC Olympics and are robust to correlations in the data.  
}


\begin{document}
\title{Simulation-Assisted Decorrelation for Resonant Anomaly Detection}

\author[1,3]{Kees Benkendorfer,} 
\author[2,3]{Luc Le Pottier,} 
\author[3]{and Benjamin Nachman}
\affiliation[1]{\normalsize Physics Department, Reed College, Portland, OR 97202, USA}
\affiliation[2]{\normalsize Physics Department, University of Michigan, Ann Arbor, MI 48109, USA}
\affiliation[3]{\normalsize Physics Division, Lawrence Berkeley National Laboratory, Berkeley, CA 94720, USA}

\emailAdd{kebenkend@reed.edu}
\emailAdd{luclepot@umich.edu}
\emailAdd{bpnachman@lbl.gov}

\maketitle
 
\section{Introduction}\label{sec:intro}

Despite compelling experimental (e.g. dark matter) and theoretical (e.g. the hierarchy problem) evidence for new phenomena at the electroweak scale, experiments at the Large Hadron Collider (LHC) have not yet discovered any physics beyond the Standard Model (BSM).  There are major search efforts across LHC experiments~\cite{atlasexoticstwiki,atlassusytwiki,atlashdbspublictwiki,cmsexoticstwiki,cmssusytwiki,cmsb2gtwiki,lhcbtwiki}, where most analyses target a particular class of BSM models.  While this work is well-motivated and continuing to improve in sensitivity (in part due to machine learning~\cite{Larkoski:2017jix,Guest:2018yhq,Abdughani:2019wuv,Radovic:2018dip}), there is also a growing need for new search strategies capable of discovery in unexpected scenarios. 

A variety of automated anomaly detection techniques using innovative machine learning methods are being proposed to cover the unexpected~\cite{DAgnolo:2018cun,Collins:2018epr,Collins:2019jip,collaboration2020dijet,DAgnolo:2019vbw,Farina:2018fyg,Heimel:2018mkt,Roy:2019jae,Cerri:2018anq,Blance:2019ibf,Hajer:2018kqm,DeSimone:2018efk,Mullin:2019mmh,1809.02977,Dillon:2019cqt,Andreassen:2020nkr,Nachman:2020lpy,Aguilar-Saavedra:2017rzt,Romao:2019dvs,Romao:2020ojy,knapp2020adversarially,1797846,1800445,Amram:2020ykb,Cheng:2020dal,Khosa:2020qrz,Thaprasop:2020mzp}.  An important subset of these proposals targets resonant new physics, where sideband methods can be used to estimate the SM background directly from data.  A key challenge facing such methods is that the machine learning classifiers must be relatively independent from the resonant feature, for otherwise artificial bumps can be formed.  Many automated decorrelation methods have been proposed to ensure that classifiers are decorrelated from particular features by construction~\cite{Louppe:2016ylz,Dolen:2016kst,Moult:2017okx,Stevens:2013dya,Shimmin:2017mfk,Bradshaw:2019ipy,ATL-PHYS-PUB-2018-014,Xia:2018kgd,Englert:2018cfo,Wunsch:2019qbo,Disco,2007.14400}, but they may not apply in all cases.  In particular, weakly supervised approaches that learn directly on the signal region cannot be simply combined with a decorrelation scheme because such an approach could degrade the performance in the presence of a signal.  A localized signal would manifest as a dependence between the resonant feature and other features for classification, so forcing independence could eliminate signal sensitivity. 

In this paper, two weakly supervised approaches are studied: Classification without Labels (\textsc{CWoLa})~\cite{Metodiev:2017vrx,Collins:2018epr,Collins:2019jip,collaboration2020dijet} and Simulation Assisted Likelihood-free Anomaly Detection (\textsc{Salad})~\cite{Andreassen:2020nkr}.  \textsc{CWoLa} is a method that does not depend on simulation and achieves signal sensitivity by comparing a signal region with nearby sideband regions in the resonance feature.  As a result, \textsc{CWoLa} is particularly sensitive to dependencies between the classification features and the resonant feature.  \textsc{Salad} uses a reweighted simulation to achieve signal sensitivity.  Since it never directly uses the sideband region, \textsc{Salad} is expected to be more robust than \textsc{CWoLa} to dependencies.  In order to recover the performance of \textsc{CWoLa} in the presence of significant dependence between the classification features and the resonant feature, a new method called simulation augmented CWoLa (SA-\textsc{CWoLa}) is introduced.  The SA-\textsc{CWoLa} approach augments the \textsc{CWoLa} loss function to penalize the classifier for learning differences between the signal region and the sideband region in simulation, which is signal-free by construction.  All of these methods will be investigated using the correlation test proposed in Ref.~\cite{Nachman:2020lpy}.

This paper is organized as follows.  Section~\ref{sec:methods} reviews the \textsc{Salad} and \textsc{CWoLa} methods and introduces the simulation augmented \textsc{CWoLa} search strategy.  Furthermore, the sideband analysis is setup in Sec.~\ref{sec:methods}.  The simulations used for illustrating the various approaches are described in Sec.~\ref{sec:sim}.  Results for the different strategies are presented in Sec.~\ref{sec:results}.  The paper ends with conclusions and outlook in Sec.~\ref{sec:conclusions}.

\section{Methods}\label{sec:methods}

For a set of features $(m,x)\in\mathbb{R}^{n+1}$, let $f:\mathbb{R}^n\rightarrow [0,1]$ be parameterized by a neural network.   The observable $m$ is special, for it is the resonance feature that should be relatively independent from $f(x)$.  The signal region (SR) is defined by an interval in $m$ and the sidebands (SB) are neighboring intervals.

All neural networks were implemented in \textsc{Keras}~\cite{keras} with the \textsc{Tensorflow} backend~\cite{tensorflow} and optimized with \textsc{Adam}~\cite{adam}.   Each network is composed of three hidden layers with 64 nodes each and use the rectified linear unit (ReLU) activation function.  The sigmoid function is used after the last layer.  Training proceeds for 10 epochs with a batch size of 200.   None of these parameters were optimized; it is likely that improved performance could be achieved with an in-situ optimization based on a validation set.

\clearpage

\subsection{Simulation Assisted Likelihood-free Anomaly Detection (SALAD)}

The \textsc{Salad} network~\cite{Andreassen:2020nkr} is optimized using the following loss:

\begin{align}
\mathcal{L}_\text{SALAD}[f]&=-\sum_{i\in\text{SR,data}}\log(f(x_i))-\sum_{i\in\text{SR,sim.}}w(x_i,m)\log(1-f(x_i))\,
\end{align}
where $w(x_i,m)=g(x_i,m)/(1-g(x_i,m))$ are a set of weights using the Classification for Tuning and Reweighting (\textsc{Dctr})~\cite{Andreassen:2019nnm} method.  The function $g$ is a parameterized classifier~\cite{Cranmer:2015bka,Baldi:2016fzo} trained to distinguish data and simulation in the sideband:

\begin{align}
\mathcal{L}[g]&=-\sum_{i\in\text{SB,data}}\log(g(x_i,m))-\sum_{i\in\text{SB,sim.}}\log(1-g(x_i,m))\,.
\end{align}
The above neural networks are optimized with binary cross entropy, but one could use other functions as well, such as the mean-squared error.  Intuitively, the idea of \textsc{Salad} is to train a classifier to distinguish data and simulation in the SR.  However, there may be significant differences between the background in data and the background simulation, so a reweighting function is learned in the sidebands that makes the simulation look more like the background in data.  

\subsection{Simulation Augmented Classification without Labels (CWoLa)}

The idea of \textsc{CWoLa}~\cite{Metodiev:2017vrx} is to construct two mixed samples of data that are each composed of two classes.  Using \textsc{CWoLa} for resonant anomaly detection~\cite{Collins:2018epr,Collins:2019jip}, one can construct the mixed samples using the SR and SB.  In the absence of signal, the SR and SB should be statistically identical and therefore the \textsc{CWoLa} classifier does not learn anything useful.  However, if there is a signal, then it can detect the presence of a difference between the SR and SB.  In practice, there are small differences between the SR and SB because there are dependencies between $m$ and $x$ and so \textsc{CWoLa} will only be able to find signals that introduce a bigger difference than already present in the background.  The \textsc{CWoLa} anomaly detection strategy was recently used in a low-dimensional application by the ATLAS experiment~\cite{collaboration2020dijet}.

We propose a modification of the usual \textsc{CWoLa} loss function in order to construct a simulation-augmented (SA) \textsc{CWoLa} classifier:

\begin{align}\nonumber
\mathcal{L}_\text{SA-CWola}[f]&=-\sum_{i\in\text{SR,data}}\log(f(x_i))-\sum_{i\in\text{SB,data}}\log(1-f(x_i))\\
\label{eq:sacwola}
&\hspace{25mm}+\lambda\left(\sum_{i\in\text{SR,sim.}}\log(f(x_i))+\sum_{i\in\text{SB,sim.}}\log(1-f(x_i))\right)\,,
\end{align}
where $\lambda > 0$ is a hyper-parameter.  The limit $\lambda\rightarrow 0$ is the usual \textsc{CWoLa} approach and for $\lambda > 0$, the classifier is penalized if it can distinguish the SR from the SB in the (background-only) simulation\footnote{One could also use the SALAD-reweighted background simulation. In practice, we found little difference between using and not using the weights as the data/sim differences were a subleading correction to the mass-dependence.  However, this may be more useful in other applications.  We thank Jesse Thaler for this interesting idea.}.  In order to help the learning process, the upper and lower sidebands are given the same total weight as each other and together, the same weight as the SR.

\subsection{Bump Hunt Analysis}
\label{sec:fitbumphunt}

In addition to quantifying performance with Receiver Operating Characteristic (ROC) curves, it is also useful to emulate a proper background estimation based on a bump hunt.  A histogram of the $m_{jj}$ spectrum, possibly after applying a threshold on one of the classifiers described above, is fit to the following parametric function:

\begin{align}
\frac{d\sigma}{dm_{jj}}=\frac{p_0\,(1-x)^{p_1}}{x^{p_x+p_3\,\log(x)}}\,,
\end{align}
where $x=m_{jj}/\sqrt{s}$ and $p_i$ are fit parameters.  This function has a long history and has also been recently used by the ATLAS and CMS collaborations (see e.g.~\cite{Sirunyan:2018xlo,Aad:2019hjw}).  Alternative non-parametric functions are also possible (such as Gaussian processes~\cite{Frate:2017mai}), but these are not needed for the demonstration considered here.  The SR is masked during the fit and then a $p$-value of the observed data is computed in the usual way.  In particular, a test statistic is formed from the profile likelihood ratio:

\begin{align}
\lambda_0 = \frac{\max_\theta p(n|\mu=0,\theta)}{\max_{\mu,\theta}p(n|\mu,\theta)},
\end{align}
where $n$ is the number of observed events in the SR and $\theta$ is a nuisance parameter from the sideband fit:

\begin{align}
p(n|\mu,\theta)=\text{Poisson}(n|b+\theta+\mu)e^{-\theta^2/2\sigma^2},
\end{align}
where $b$ and $\sigma$ are the number of events and uncertainty from the sideband fit, respectively.  The test statistic itself is $q_0=-2\log(\lambda_0)$ when the extracted signal strength $(\mu,\theta)=\text{argmax}_{\mu',\theta'} p(n|\mu',\theta')$ is $\mu>0$ and 0 otherwise.  Asymptotic formulae from Wald and Wilks then give the significance $Z=\sqrt{q_0}$~\cite{wilks1938,10.2307/1990256,Cowan:2010js}.

In practice, one would scan the signal region across the $m_{jj}$ spectrum.  In this analysis, we will focus on a single region with or without signal injected.  The signal region is defined by $m_{jj}\in[3.3,3.7]$ TeV and the sideband for \textsc{CWoLa} training is defined as $m_{jj}\in[3.1,3.3]\cup[3.7,3.9]$ TeV.  Long sidebands extended by 300 GeV in either direction are used to train the \textsc{Salad} reweighting function.  The background fit is performed between 2.6 and 5 TeV using 30 equally-spaced bins.

\section{Simulation}\label{sec:sim}

The simulations used for this study were produced for the LHC Olympics 2020 community challenge~\cite{gregor_kasieczka_2019_2629073}.  In particular, the background process is composed of generic dijet events with a requirement for at least one such jet with $p_T>1.3$ TeV.  Signal events are $W'\rightarrow XY$ for $m_{W'}=3.5$ TeV and hypothetical particles $X$ and $Y$ of mass 500 and 100 GeV, each decaying into pairs of quarks.  Due to the mass hierarchy between the $W'$ boson and its decay products, the final state is characterized by two large-radius jets with two-prong substructure.  The background and signal are simulated using \textsc{Pythia}~8~\cite{Sjostrand:2006za,Sjostrand:2007gs} and an alternative background sample is simulated using \textsc{Herwig}++~\cite{Bahr:2008pv}.  A detector simulation is performed with Delphes~3.4.1~\cite{deFavereau:2013fsa,Mertens:2015kba,Selvaggi:2014mya} using the default CMS detector card.  Particle flow objects are the input to jet clustering, implemented using \textsc{Fastjet}~\cite{Cacciari:2011ma,Cacciari:2005hq} and the anti-$k_t$ algorithm~\cite{Cacciari:2008gp} using $R=1.0$ for the radius parameter.   In what follows, \textsc{Pythia} will play the role of `data' and the \textsc{Herwig} sample will be used as the `simulation'.  There are one million events for both background samples, corresponding to an integrated luminosity of about 100 fb$^{-1}$.  In order to simplify the analysis, the dataset is divided in half for training and testing.  More complicated procedures based on $k$-folding to use the entire dataset for both training and testing are also possible, but are not considered here~\cite{Collins:2018epr,Collins:2019jip}. 

Both the \textsc{CWoLa} and \textsc{Salad} methods have been demonstrated on the unmodified LHC Olympics dataset.  Following Ref.~\cite{Nachman:2020lpy}, the dependence between the jet masses and $m_{jj}$ is artificially strengthened by redefining $m_{j}\mapsto m_{j}+\alpha\, m_{jj}$ for $\alpha=0.1$.  As shown in Ref.~\cite{Nachman:2020lpy}, this shift is sufficient to reduce the efficacy of the unmodified \textsc{CWoLa} method.

\begin{figure}[h!]
\centering
\includegraphics[height=0.39\textwidth]{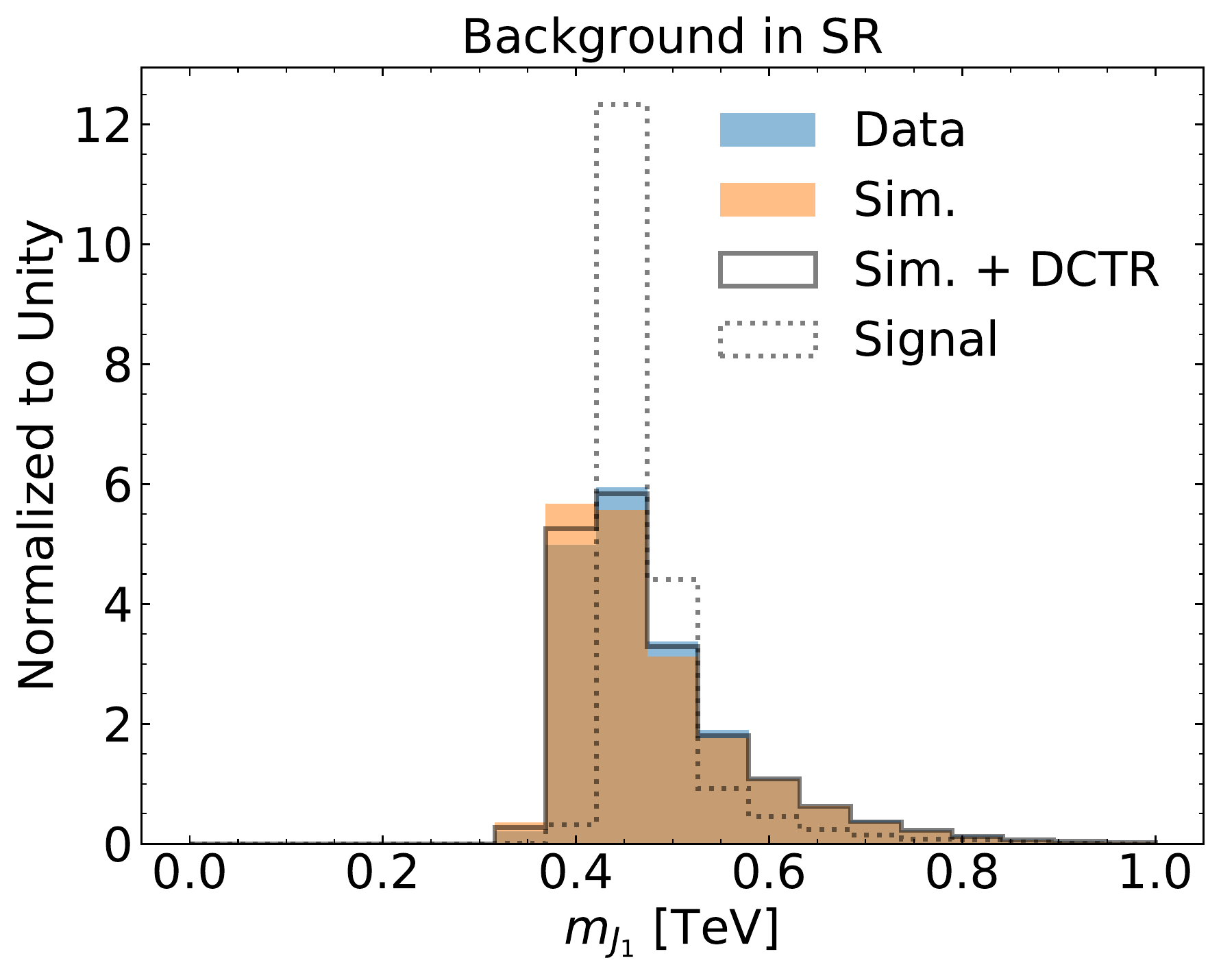}\includegraphics[height=0.39\textwidth]{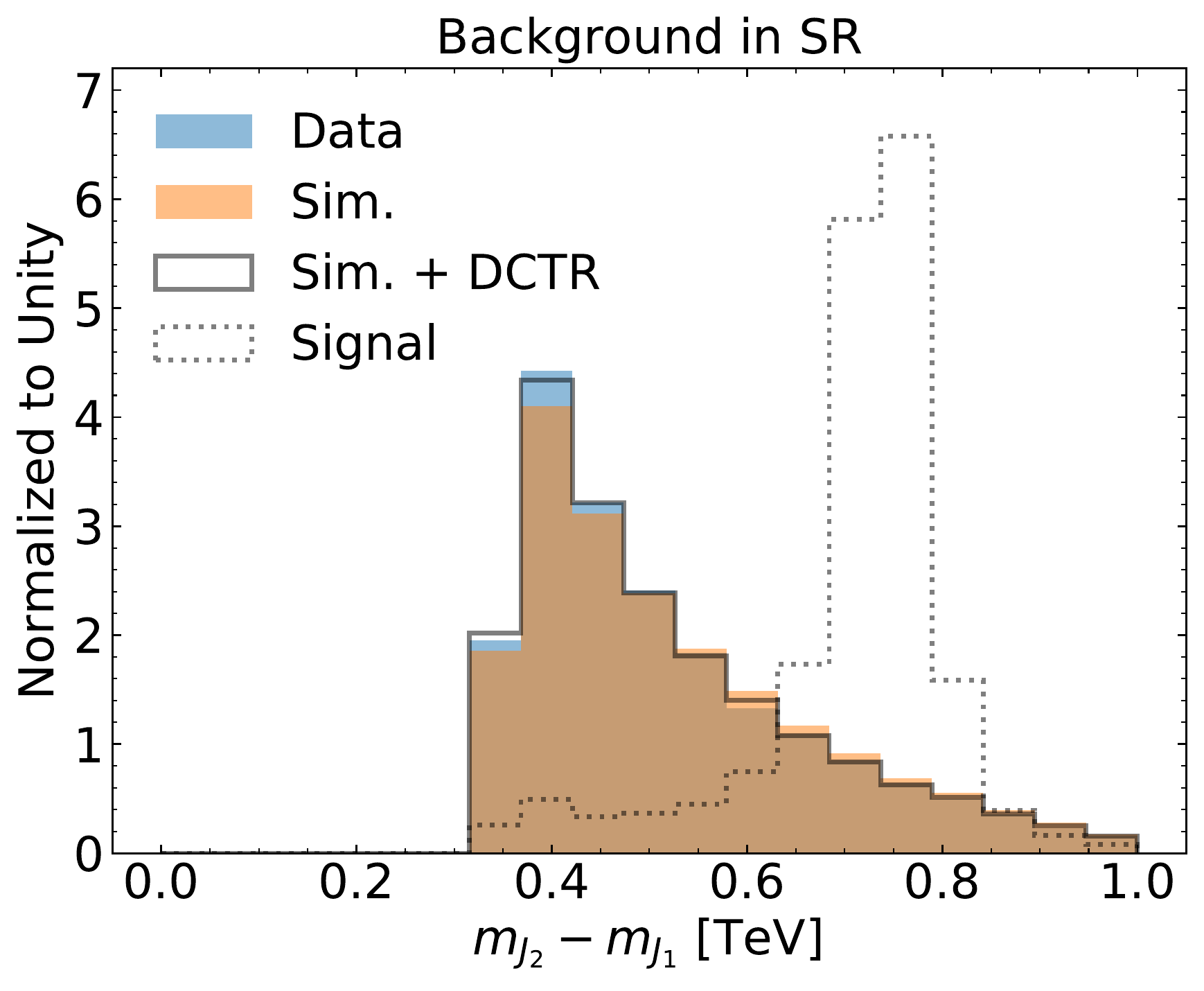}\\
\includegraphics[height=0.39\textwidth]{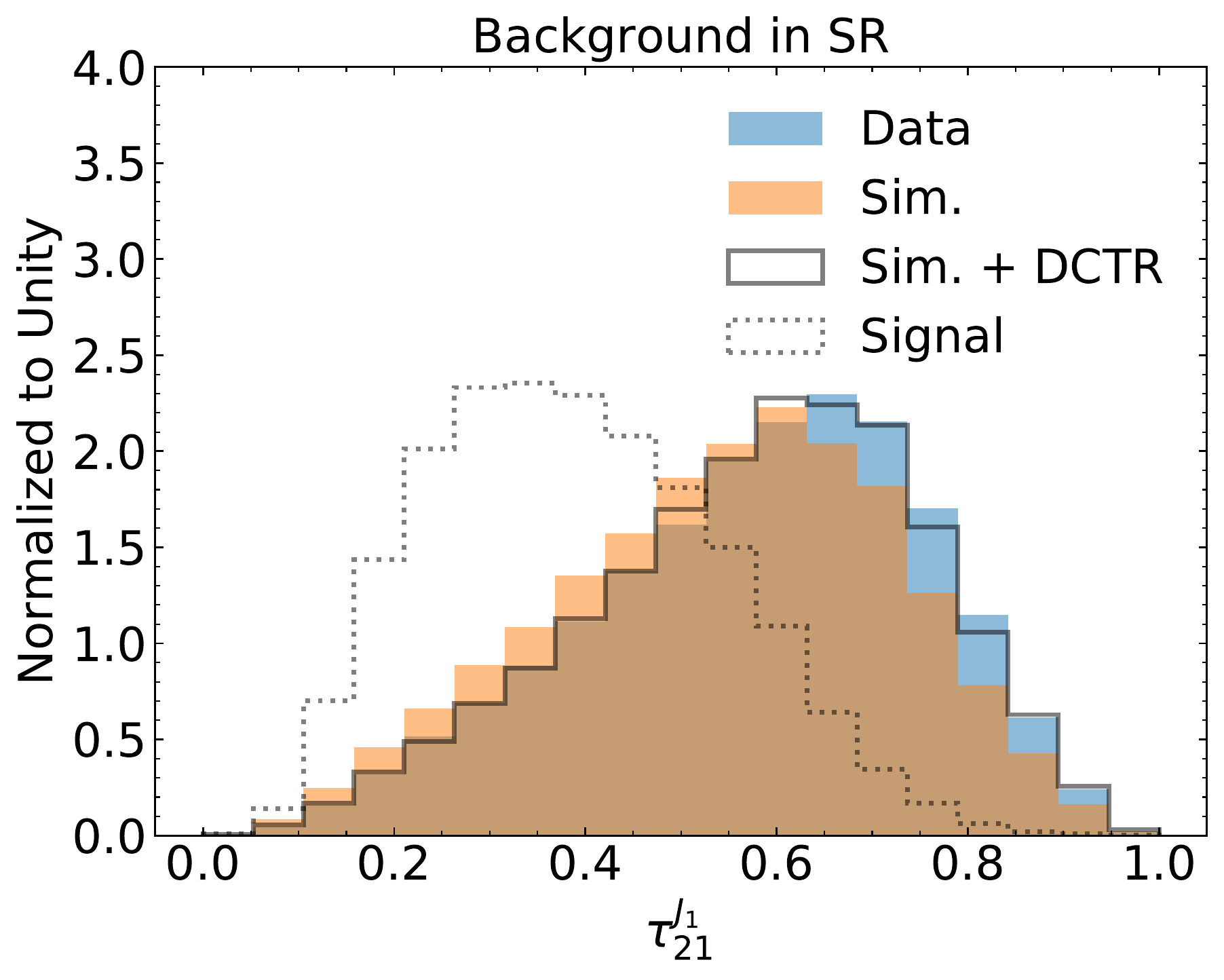}\includegraphics[height=0.39\textwidth]{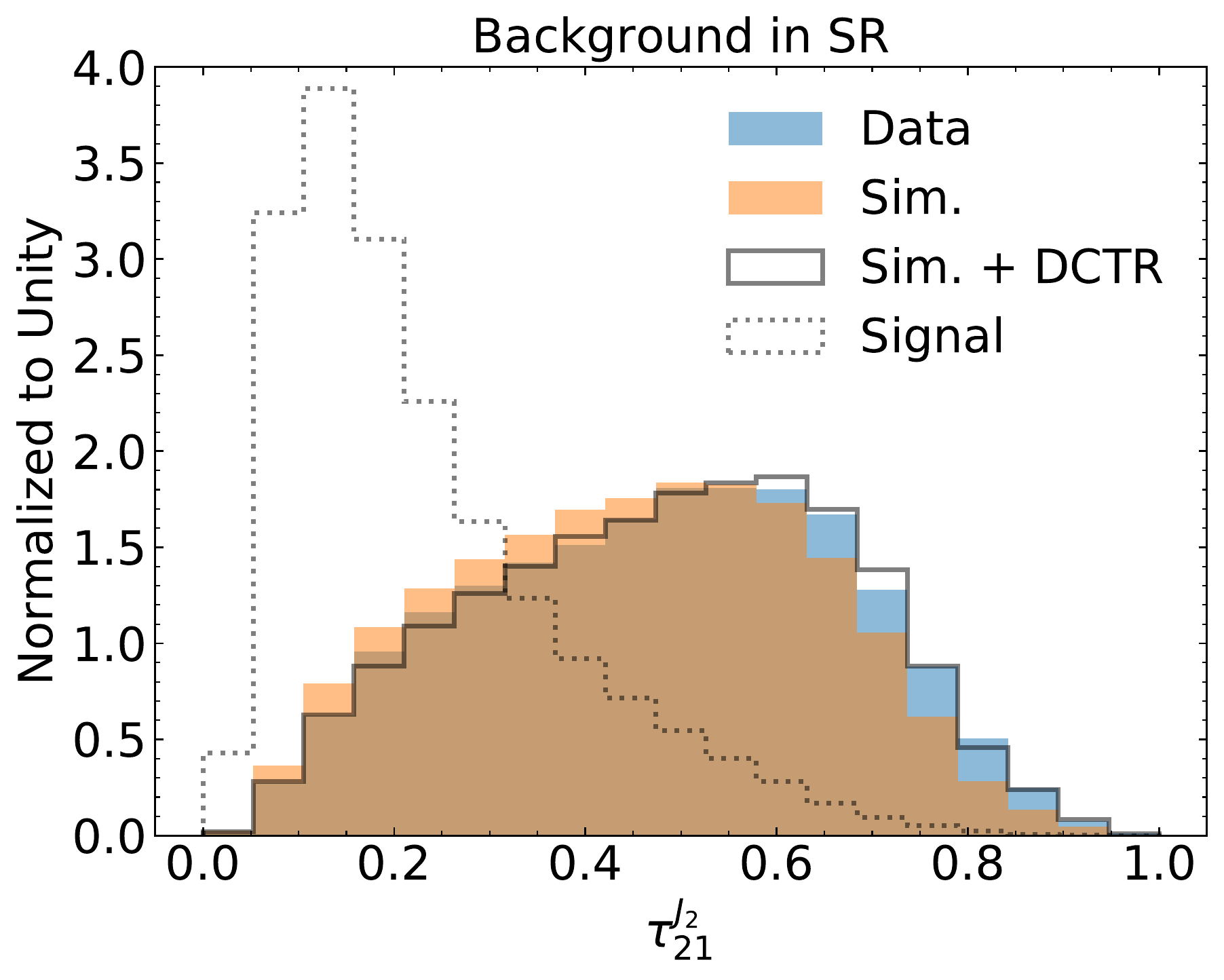}
\caption{Left: the jet mass and $\tau_{21}$ of the jet with a smaller mass.  Right: the difference between the heavier and lighter jet masses and $\tau_{21}$ of the heavier jet.  In addition to showing the data, simulation, and signal, the histogram labeled `Sim.+DCTR' is the simulation with weights derived from a parameterized reweighting function trained on long sidebands.}
\label{fig:features}
\end{figure}

In addition to the dijet invariant mass, four features are used for the anomaly detection: the invariant mass of the lighter jet, the mass difference of the leading two jets, and the $\tau_{21}$~\cite{Thaler:2011gf,Thaler:2010tr} of the leading two jets.  The $N$-subjettiness $\tau_{21}$ quantifies the extent to which a jet is characterized by two subjets or one subjet.  Histograms of the four input features for the background are shown in Fig.~\ref{fig:features}.  The signal jet masses are localized at the $X$ and $Y$ masses (shifted by $\alpha\, m_{W'}$) and the $\tau_{21}$ are shifted to lower values, indicating two-pronginess.  In addition to presenting the data and simulation histograms, Fig.~\ref{fig:features} also shows the reweighted background simulation using parameterized weights learned from a long sideband.

\clearpage

\section{Results}\label{sec:results}

\begin{figure}[h!]
\centering
\includegraphics[height=0.36\textwidth]{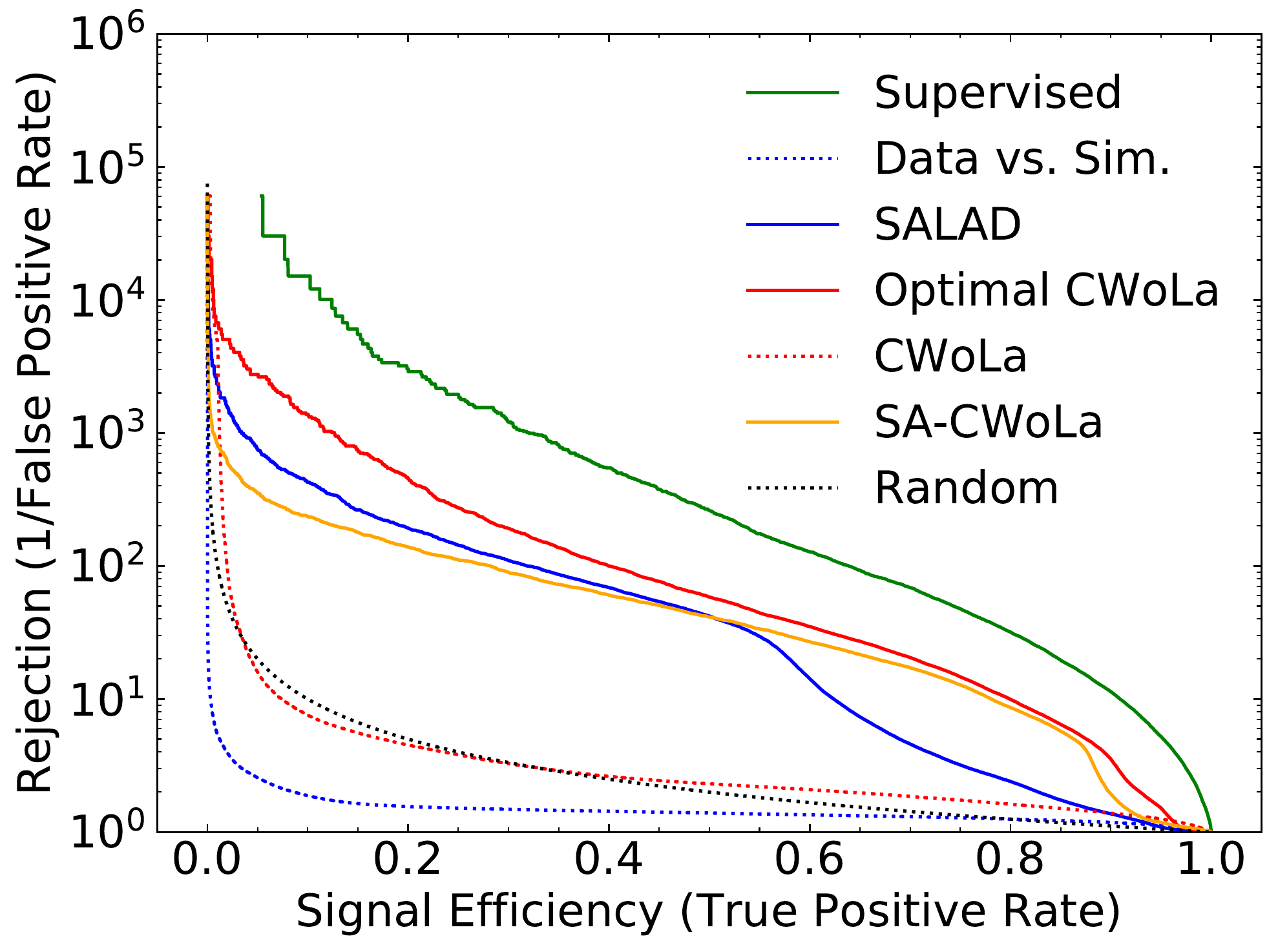}\includegraphics[height=0.36\textwidth]{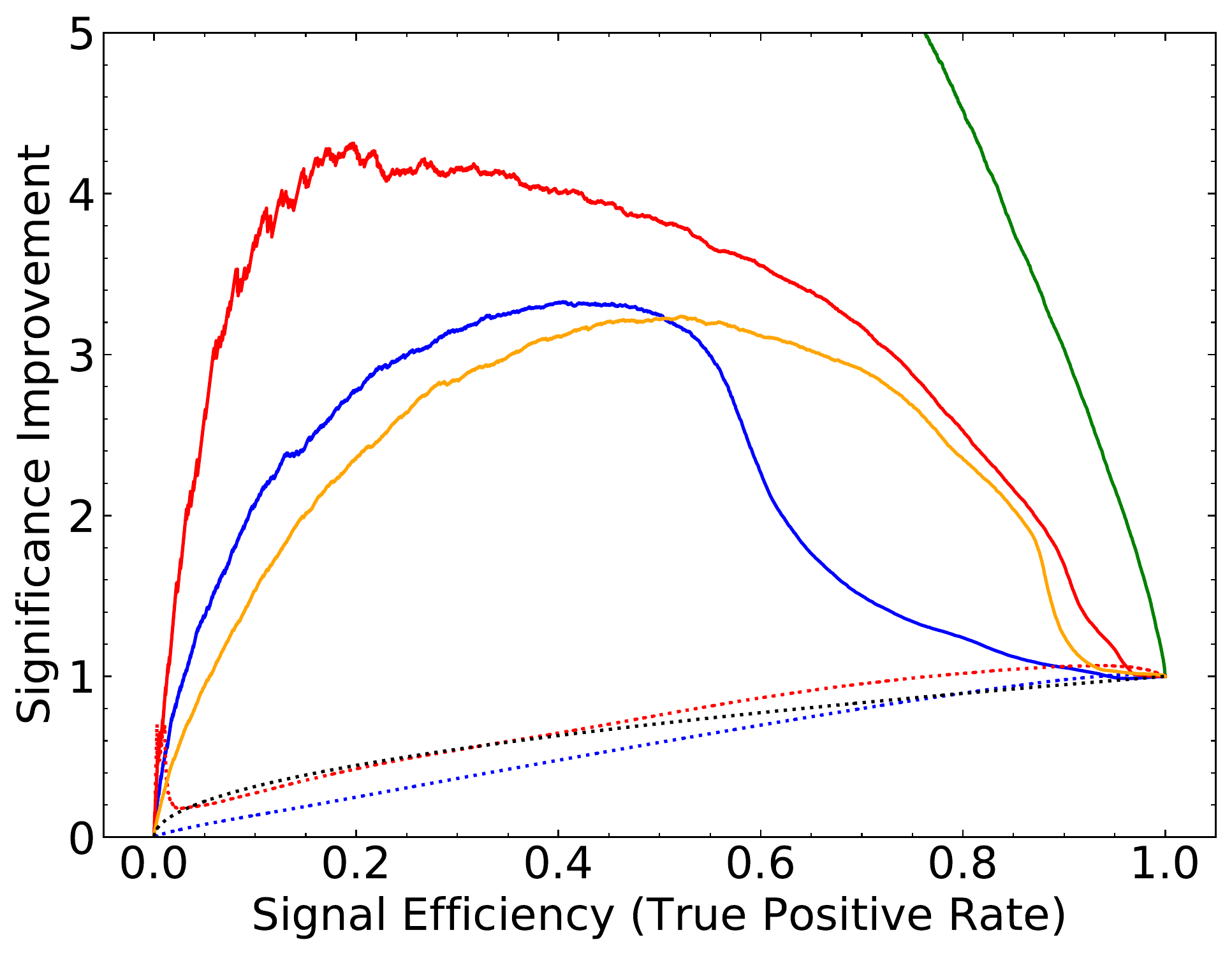}
\caption{A Receiver Operating Characteristic (ROC) curve (left) and significance improvement curve (right) for various anomaly detection methods described in the text.   The significance improvement is defined as the ratio of the signal efficiency to the square root of the background efficiency.  A significance improvement of 2 means that the initial significance would be amplified by about a factor of two after employing the anomaly detection strategy.  The supervised line is unachievable unless there is no mismodeling and one designed a search for the specific $W'$ signal used in this paper.  The curve labeled `Random' corresponds to equal efficiency for signal and background.}
\label{fig:roc}
\end{figure}

As a benchmark, 1500 signal events corresponding to a fitted significance of about $2\sigma$ is injected into the data for training.  For evaluation, the entire signal sample (except for the small number of injected events) is used.  Figure~\ref{fig:roc} shows the performance of various configurations.    The fully supervised classifier uses high statistics signal and background samples in the SR with full label information.  Since the data are not labeled, this is not achievable in practice.  A solid red line labeled `Optimal \textsc{CWoLa}' corresponds to a classifier trained using two mixed samples, one composed of pure background in the single region and the other composed of mostly background (independent from the first sample) in the SR with the 1500 signal events.  This is optimal in the sense that it removes the effect from phase space differences between the SR and SB for the background.  The Optimal \textsc{CWoLa} line is far below the fully supervised classifier because the neural network needs to identify a small difference between the mixed samples over the natural statistical fluctuations in both sets.  The actual \textsc{CWoLa} method is shown with a dotted red line.  By construction, there is a significant difference between the phase space of the SR and SB and so the classifier is unable to identify the signal.  At low efficiency, the \textsc{CWoLa} classifier actually anti-tags because the SR-SB differences are such that the signal is more SB-like then SR-like.  Despite this drop in performance, the simulation augmenting modification (solid orange) with $\lambda=0.5$ nearly recovers the full performance of \textsc{CWoLa}.  

For comparison, a classifier trained using simulation directly is also presented in Figure~\ref{fig:roc}.  The line labeled `Data vs. Sim.' directly trains a classifier to distinguish the data and simulation in the SR without reweighting.  Due to the differences between the background in data and the simulated background, this classifier is not effective.  In fact, the signal is more like the background simulation than the data background and so the classifier is worse than random (preferentially removes signal).  The performance is significantly improved by adding in the parameterized reweighting, as advocated by Ref.~\cite{Andreassen:2020nkr}.  With this reweighting, the \textsc{Salad} classifier is significantly better than random and is comparable to SA-\textsc{CWoLa}.  The Optimal \textsc{CWoLa} line also serves as the upper bound in performance for \textsc{Salad} because it corresponds to the case where the background simulation is statistical identical to the background in data.

\begin{figure}[h!]
\centering
\includegraphics[height=0.36\textwidth]{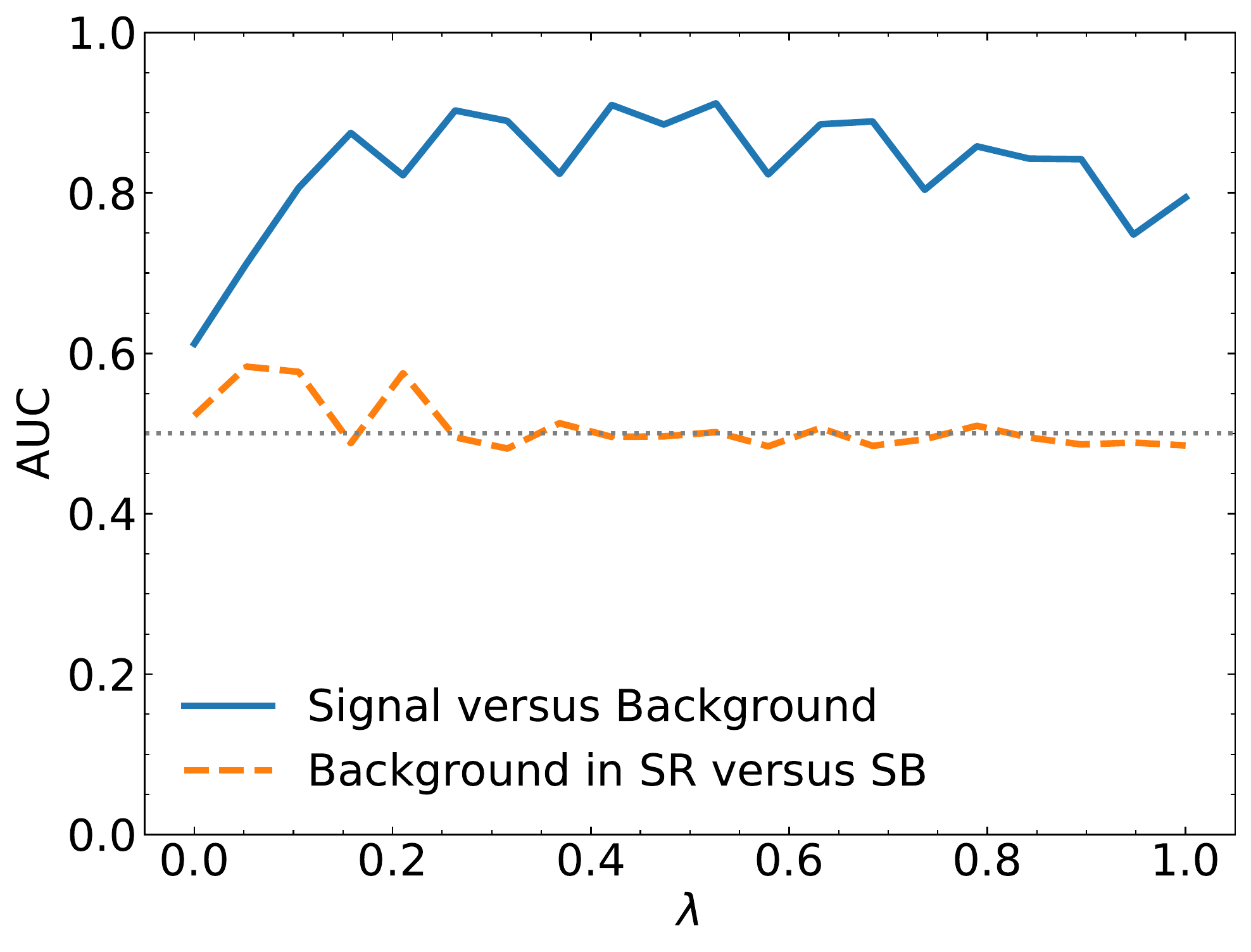}\includegraphics[height=0.36\textwidth]{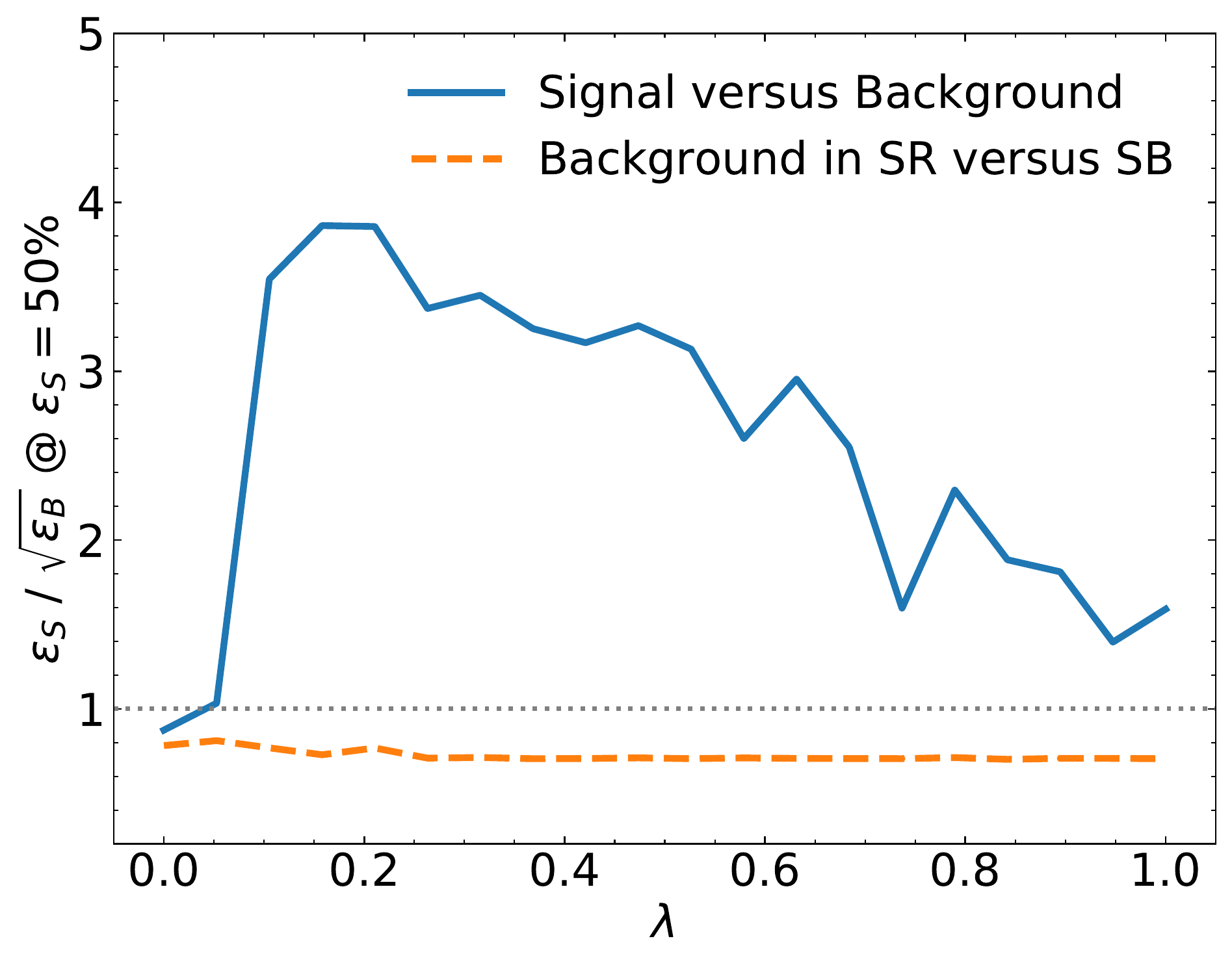}
\caption{The Area Under the ROC Curve (AUC) (left) and significance improvement at 50\% signal efficiency (right) using the SA-\textsc{CWoLa} method for a scan in the hyperparameter $\lambda$ introduced in Eq.~\ref{eq:sacwola}.  When $\lambda=0$, SA-\textsc{CWoLa} is the same as the original \textsc{CWoLa} method.  For comparison, the performance of the classifier for distinguishing signal and background is shown in blue and the performance for distinguishing the SR and SB is shown in orange.  Ideally, the latter would have an AUC of 0.5.}
\label{fig:lambdascan}
\end{figure}

The SA-\textsc{CWoLa} method has one free parameter that must be tuned.   Figure~\ref{fig:lambdascan} quantifies the performance of the SA-\textsc{CWoLa} classifier as a function of $\lambda$.   The performance of SA-\textsc{CWoLa} is strong and relatively stable for $0.3 < \lambda < 0.6$.   For $\lambda \gtrsim 0.2$, the classifier is effectively blinded to differences between the SR and SB as illustrated in the orange lines in Fig.~\ref{fig:lambdascan} approaching 0.5 in the left plot.

\begin{figure}[h!]
\centering
\includegraphics[height=0.38\textwidth]{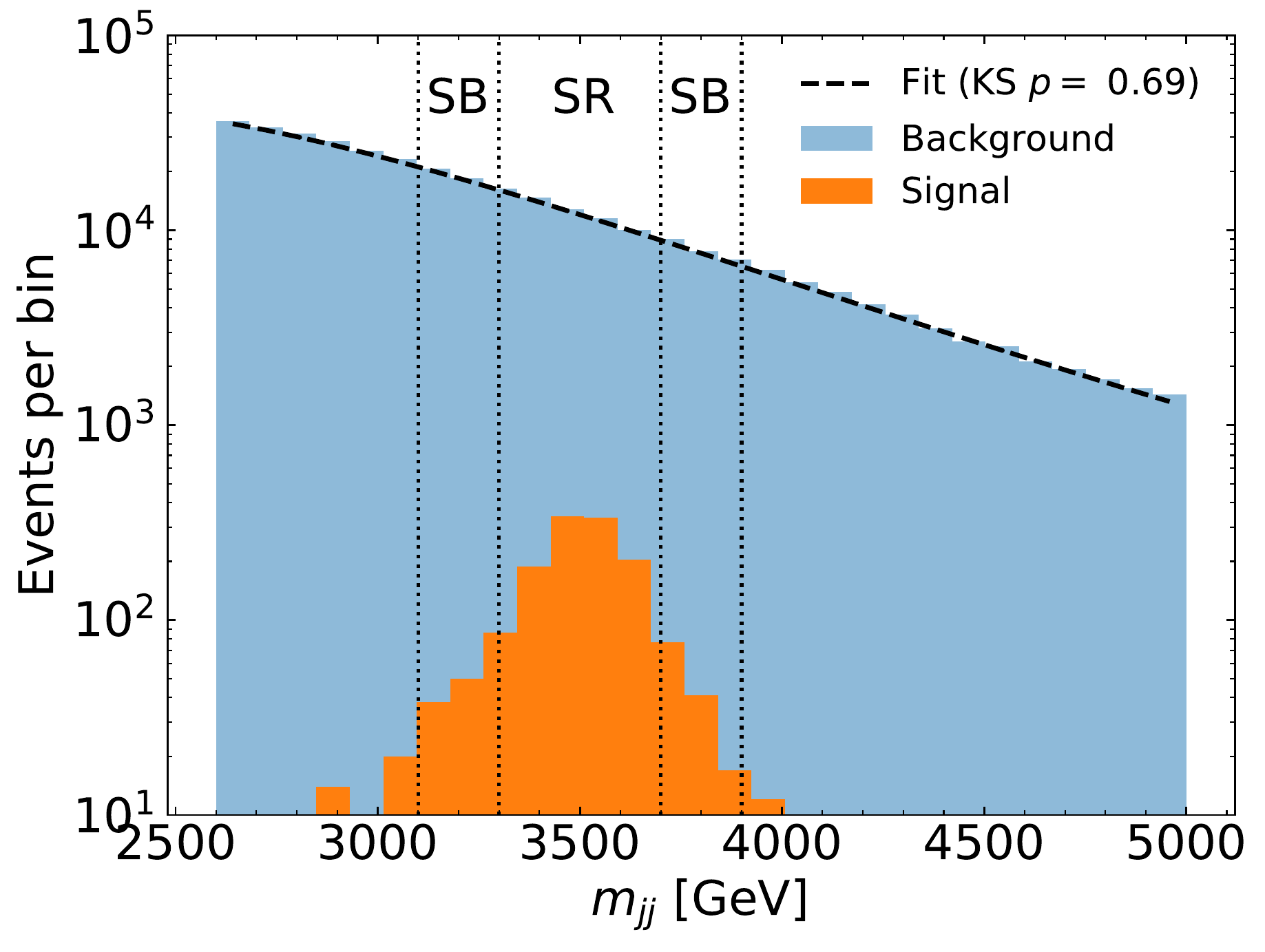}
\caption{A fit to the $m_{jj}$ distribution in the background-only case with no selection on any neural networks.   The 1500 signal events used for training is super-imposed for illustration.  Vertical dashed lines indicate the SR and SB regions used for training.  A Kolmogorov-Smirnov (KS) test using only bins outside of the SR yields a $p$-value of 0.69.}
\label{fig:mjjfit}
\end{figure}

\begin{figure}[h!]
\centering
\includegraphics[width=0.9\textwidth]{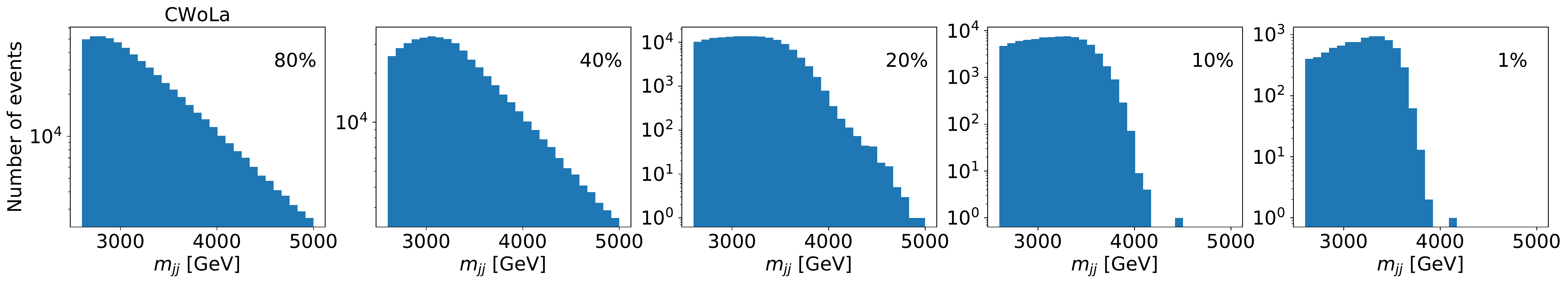}\\\includegraphics[width=0.9\textwidth]{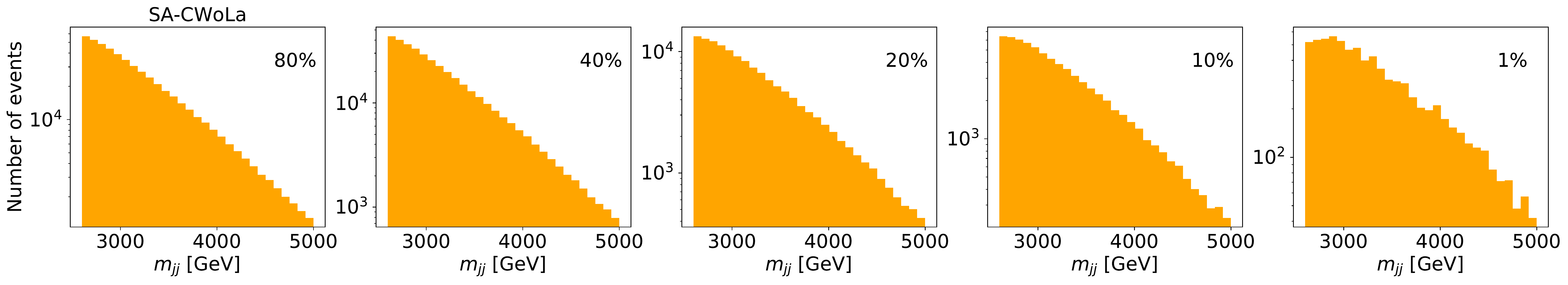}
\caption{Histograms of $m_{jj}$ for \textsc{CWoLa} (top) and \textsc{SA-CWoLa} (bottom) for various thresholds on the classifiers in the background-only case.}
\label{fig:mjj}
\end{figure}

\begin{figure}[h!]
\centering
\includegraphics[height=0.36\textwidth]{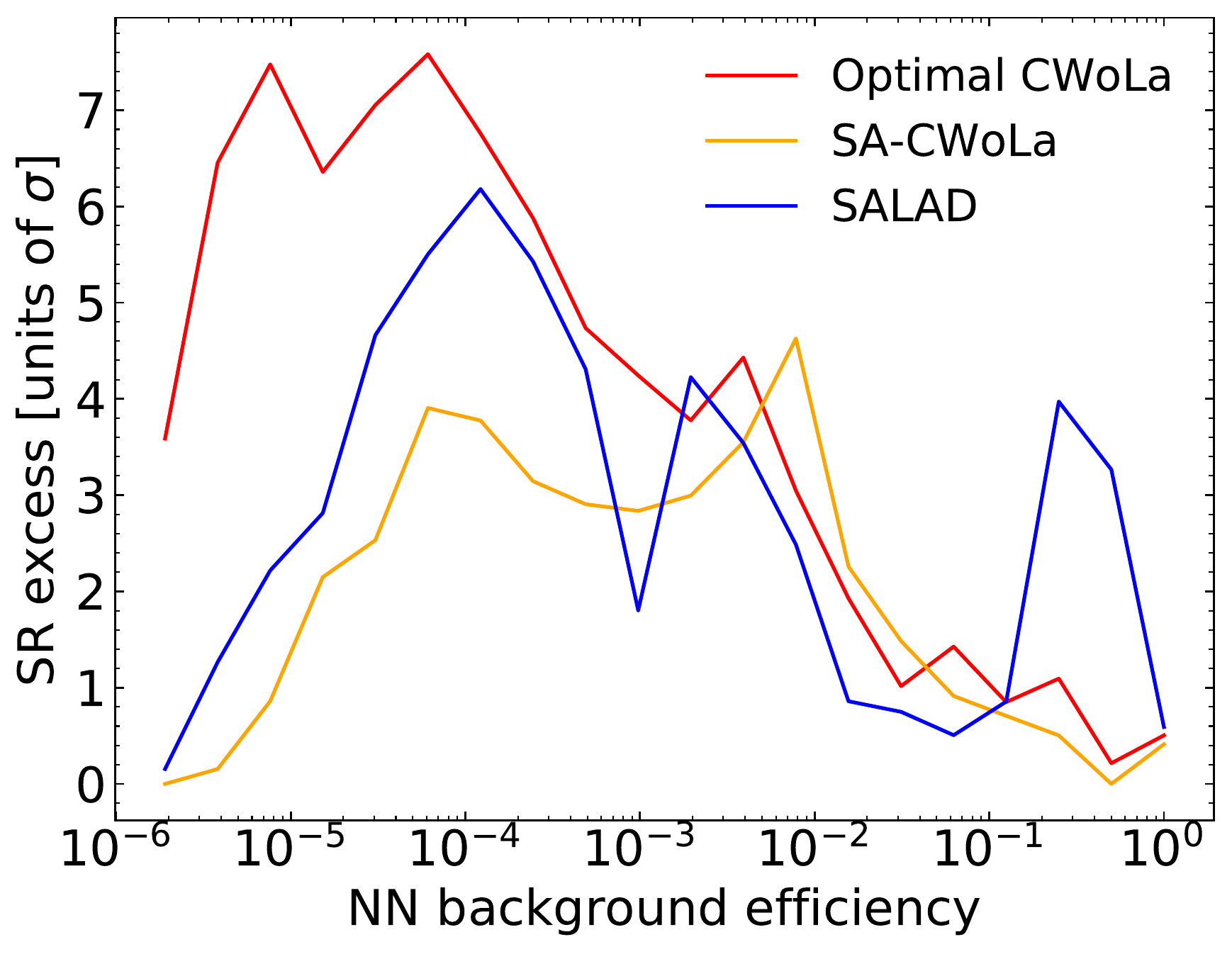}\includegraphics[height=0.36\textwidth]{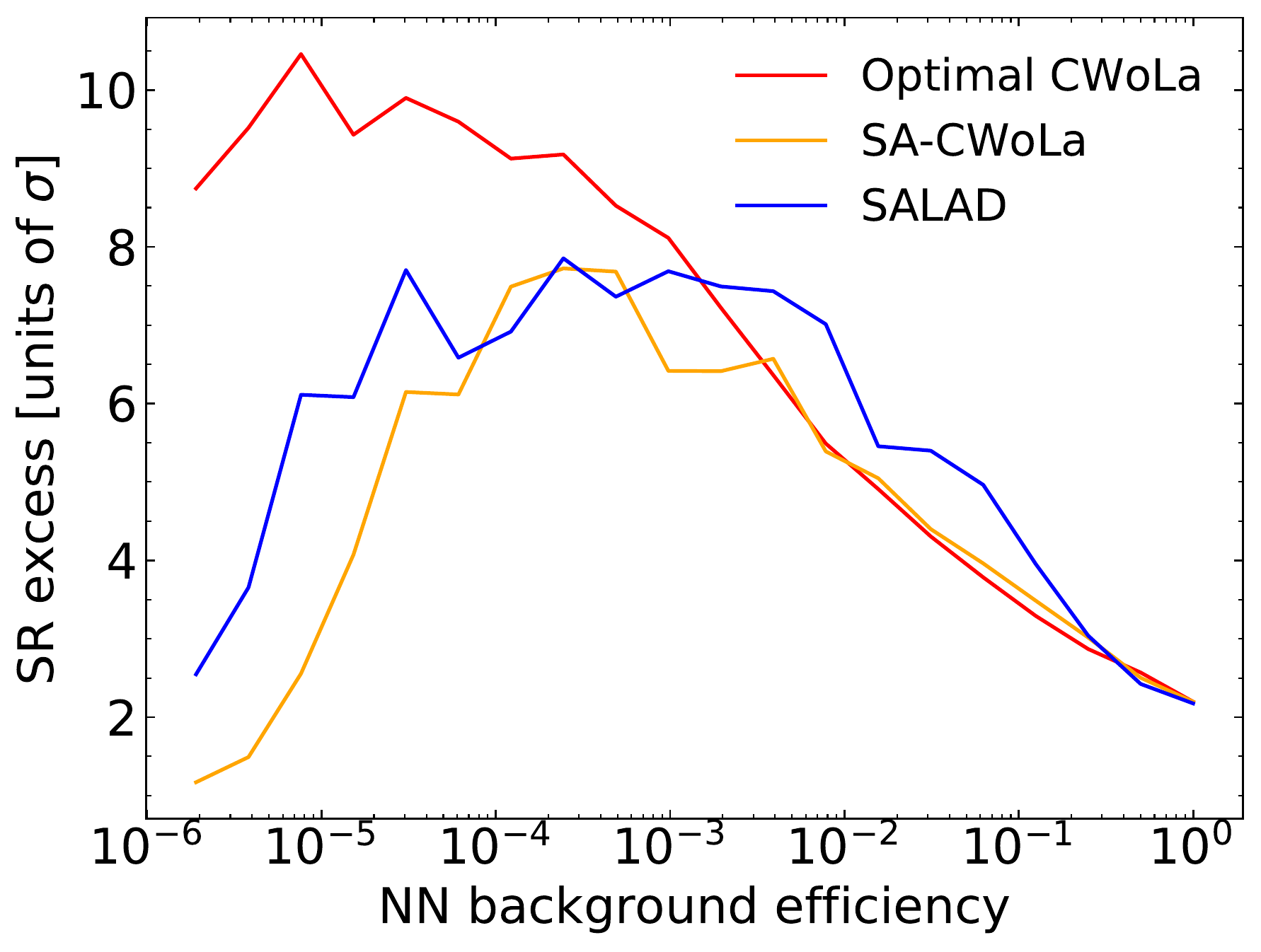}
\caption{Fit excess with signal injected using the statistical procedure described in Sec.~\ref{sec:fitbumphunt}.  There is a small local deficit in the simulation.  The left plot shows the fitted excess without modifying the background while the right plot corrects for the initial deficit by subtracting the residuals of the background-only fit before performing the signal+background fit.  In the latter case, the significances are still not $S/\sqrt{B}$ due to the uncertainty from the sideband fit.}
\label{fig:fitwithsignal}
\end{figure}

\begin{figure}[h!]
\centering
\includegraphics[height=0.36\textwidth]{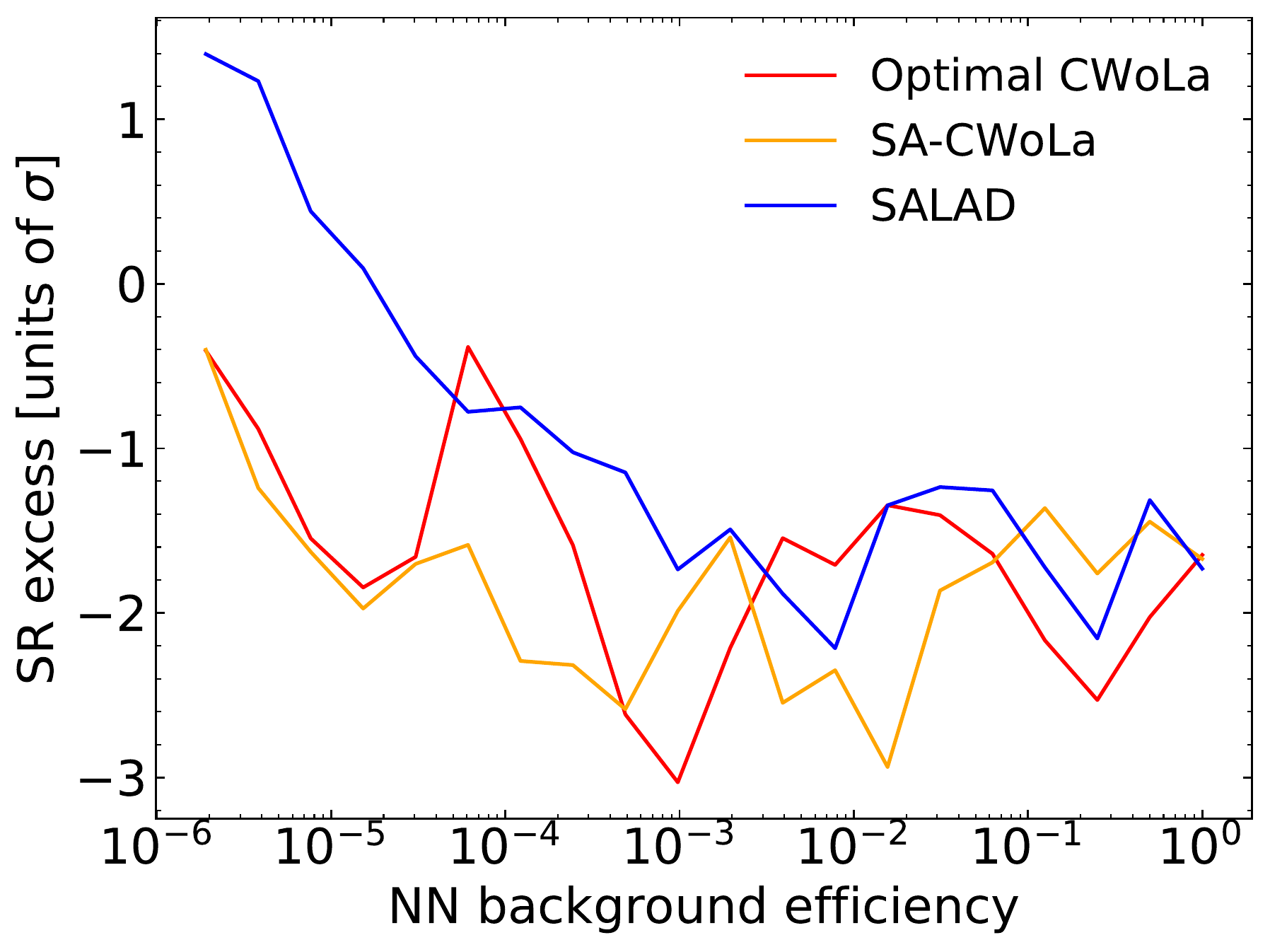}\includegraphics[height=0.36\textwidth]{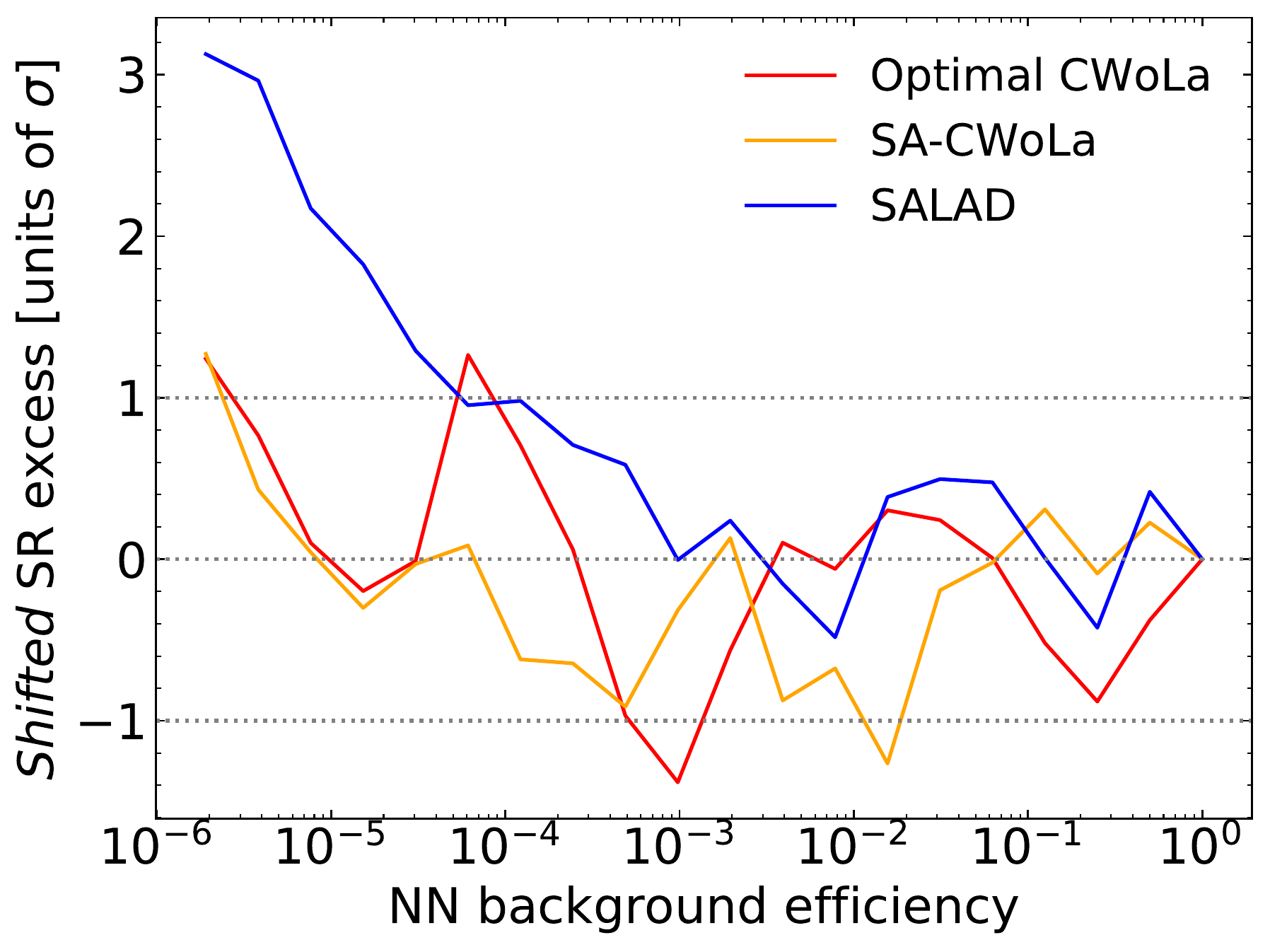}
\caption{Fit excess without signal injected using the statistical procedure described in Sec.~\ref{sec:fitbumphunt}.  Without any signal injected, there is a small $(\sim 1.5\sigma)$ deficit in the simulation. The right plot shifts the curves so that the 100\% efficiency point corresponds to $0\sigma$.}
\label{fig:fitwithoutsignal}
\end{figure}

While ROC and significance improvement curves are effective for quantifying performance, they do not communicate the complete story because they ignore the impact of background estimation.  Figures~\ref{fig:fitwithsignal} and~\ref{fig:fitwithoutsignal} show the results of the sideband fit and statistical test (See Sec.~\ref{sec:fitbumphunt}).  The fit quality is excellent when considering all bins (see Fig.~\ref{fig:mjjfit}), but there happens to be a small local deficit in the SR.   The right plot of Fig.~\ref{fig:fitwithsignal} removes this effect by subtracting the fitted residuals in the background-only case for each value of the NN background efficiency.  The spectra after applying the nominal CWoLa classifier cannot be fit to the same shape and are thus not included - see Fig.~\ref{fig:mjj}.

\clearpage

\section{Conclusions} \label{sec:conclusions}

This paper has investigated the impact of dependencies between $m_{jj}$ and classification features for the resonant anomaly detection methods \textsc{Salad} and \textsc{CWoLa}.   A new simulation-augmented approach has been proposed to remedy challenges with the \textsc{CWoLa} method.  This modification is shown to completely recover the performance of \textsc{CWoLa} from the ideal case where dependences are ignored in the training.  In both the \textsc{Salad} and SA-\textsc{CWoLa} methods, background-only simulations provide a critical tool for mitigating the sensitivity of the classifiers on dependences between the resonant feature and the classifier features.  

These weakly supervised methods are particularly promising, but they are not the only recently-proposed machine-learning based anomaly detection methods.  In particular, unsupervised methods also have great potential.  The Anomaly Detection with Density Estimation (\textsc{Anode})~\cite{Nachman:2020lpy} does not use simulation at all and has been shown to be relatively robust to dependencies between  the resonant feature and the classifier features.  Additionally, autoencoder methods have been combined with explicit decorrelation to build in robustness to such dependencies~\cite{Heimel:2018mkt}.  

Each of these unsupervised and semisupervised methods have advantages and weaknesses and it is likely that multiple approaches will be required to achieve broad sensitivity to BSM physics.  Therefore, it is critical to study the sensitivity of each technique to dependencies and propose modifications where possible to build robustness.  This paper is an important step in the decorrelation program for automated anomaly detection with machine learning.  Tools like the ones proposed here may empower higher-dimensional versions of the existing ATLAS search~\cite{collaboration2020dijet} as well as other related searches by other experiments in the near future.

\section*{Code and Data}

The code for this paper can be found at \url{https://github.com/bnachman/DCTRHunting} and the simulated data are available from the LHC Olympics~\cite{gregor_kasieczka_2019_2629073}.

\section*{\label{sec::acknowledgments}Acknowledgments}

BN would like to thank Jack Collins for useful discussions and Jesse Thaler for helpful feedback on the manuscript.  This work was supported by the Department of Energy, Office of Science under contract number DE-AC02-05CH11231.  KB was supported in part by NSF PHY REU Grant 1949923.  LLP was supported in part by the U.S. Department of Energy, Office of Science, Office of Workforce Development for Teachers and Scientists (WDTS) under the Science Undergraduate Laboratory Internships Program (SULI). BN would like to thank NVIDIA for providing Volta GPUs for neural network training.

\bibliography{refs,HEPML,myrefs}
\bibliographystyle{utphys}

\clearpage

\appendix

\end{document}